\documentclass[aps,pre,floatfix,12pt,showkeys]{revtex4-1}
\usepackage{amsfonts}
\usepackage[T2A]{fontenc}
\usepackage[utf8]{inputenc}
\usepackage{amsmath}
\usepackage{amssymb}
\usepackage[english]{babel}
\usepackage{graphicx}
\usepackage{mathtools}

\begin{document}

\title{Damped oscillations of the probability of random events followed by absolute refractory period: exact analytical results}

\author{A.V. Paraskevov$^{1,2}$, A.S. Minkin$^{2}$}

\affiliation{$^1$Institute for Information Transmission Problems, 127051 Moscow, Russia\\$^2$National Research Centre "Kurchatov Institute", 123182 Moscow, Russia}

\begin{abstract}
\begin{center}
\textbf{Abstract}
\end{center}
\noindent There are numerous examples of natural and artificial processes that represent stochastic sequences of events followed by an absolute refractory period during which the occurrence of a subsequent event is impossible. In the simplest case of a generalized Bernoulli scheme for uniform random events followed by the absolute refractory period, the event probability as a function of time can exhibit damped transient oscillations. Using stochastically-spiking point neuron as a model example, we present an exact and compact analytical description for the oscillations without invoking the standard renewal theory. The resulting formulas stand out for their relative simplicity, allowing one to analytically obtain the amplitude damping of the 2nd and 3rd peaks of the event probability.

\bigskip


\end{abstract}

\maketitle

\noindent \textbf{1. Introduction}


\noindent Natural and technical events are often followed by a refractory period: after the event has happened, repeating similar events for some time is either unlikely (relative refractory period) or impossible (absolute refractory period). A characteristic natural example is the refractory nature of the neuron ability to generate electrical pulses, "spikes". In turn, a typical technical example is the so-called dead time for some type of detectors, particularly photodetectors: the dead time is a fixed period after the detector triggering during which it becomes inoperative. Both the neuronal refractoriness and the detector dead time make an essential impact on the statistical distribution of counted events. Assuming that instant identical events occur randomly and that each event is followed by the same absolute refractory period $\tau_{ref}$, it is intuitively clear that average interval $T$ between the events should be a sum, $T=\tau_{ref}+T_{0}$, where $T_{0}$ is the average interval if the refractoriness would vanish. A more nontrivial consequence of the refractory period is damped oscillations of the event occurrence probability as a function of time \cite{Perk67,WCBJ72,Kin06}. These are strongly pronounced at $T_{0}\lesssim \tau_{ref}$. Such transient oscillations arise due to (i) a specific initial condition and (ii) the fact that, in the presence of refractoriness, the random instant events ordered in time are no longer independent from each other. In particular, the probability of a subsequent event depends on the time elapsed from the preceding one. Random point processes of this kind are called renewal processes and are a subject of study of the renewal theory \cite{Cox62}. This theory has a method of getting an explicit analytical description for the damped oscillations of the event probability by finding the so-called renewal density \cite{Perk67,Cox62}. For instance, it has been done repeatedly for the classic example of the Poisson process modulated by the dead time or absolute refractory period \cite{Malm47,Ric66,Muller73,Muller74,Cant75,John83,John86,Pom99,Picib08,PRE2010,JCN2012,NC2018}. However, the resulting analytical formula for the time-dependent probability of the event is quite bulky, and that complicates substantially getting further analytical findings.

In this paper, for the Bernoulli process modulated by absolute refractoriness, we give a compact analytical description of the damped oscillations without invoking the renewal theory. The description is presented in four equivalent forms: three variants of a recurrence formula and the explicit formula as a series. These are quantitatively consistent with numerical simulations and known results of the renewal theory. One of the recurrence formula variants is particularly simple, enabling accurate analytical calculation of the damping coefficients. Notably, these are quite robust against changing the values of the model parameters.


\noindent \textbf{2. Model description and problem statement}


\noindent For certainty, we consider a model neuron that stochastically emits spikes, each of which is followed by the absolute refractory period. In particular, the neuron can spontaneously emit a spike with probability $p_{s}$ per unit time (i.e., in a given elementary time interval $\triangle t$) so that after the spike emission the neuron becomes temporarily inactive: it cannot emit spikes during the refractory period $\tau_{ref}=n_{ref}\triangle t$, where $n_{ref}$ is a positive integer. The mean occurrence rate of spiking events in the absence of refractoriness is given by $\nu_{0}=1/T_{0}=p_{s}/\triangle t$. At $\tau_{ref}\neq 0$, from the equality $T=\tau_{ref}+T_{0}$, for the mean rate $\nu=1/T$ one gets $\nu=\nu_{0}/(1+\tau_{ref}\nu_{0})$. In fact, this mean rate is equal to the ratio of the total number of spikes to the observation time $T_{obs}$, given that $T_{obs} >> \tau_{ref}$, $T_{0}$. It is also useful to introduce an asymptotic value for the average probability of spike generation at each step, $\bar{p}_{s}=\nu \triangle t=p_{s}/(1+n_{ref}p_{s})$, such that, by the analogy with the formula for $\nu_{0}$,
\begin{equation}
\nu=\bar{p}_{s}/\triangle t=p_{s}/(\triangle t+p_{s}\tau_{ref}).\label{nu}
\end{equation}
The algorithm for simulating the neuron's dynamics is quite simple and as follows. Dividing the observation interval $T_{obs}$ by $N$ equal steps $\triangle t$, $T_{obs}=N\triangle t$, these time steps are numbered by a sequence of natural numbers starting with 1. At each step, a random number $\xi$, uniformly distributed from zero to one, is  generated and compared with the given probability $p_{s}$ of generating a spike. If $\xi\leq p_{s}$, it is assumed that spike has been generated at this time step. After the event, the neuron cannot emit a next spike during refractory period $\tau_{ref}=n_{ref}\triangle t$. The event of spike generation at the arbitrary $k$-th step is further denoted by $A_{k}$. For definiteness, the initial state of the neuron is chosen as a moment when the neuron has just left the refractory state after emitting a spike.
\begin{figure}[!t]
\centering
\includegraphics[width=0.8\textwidth]{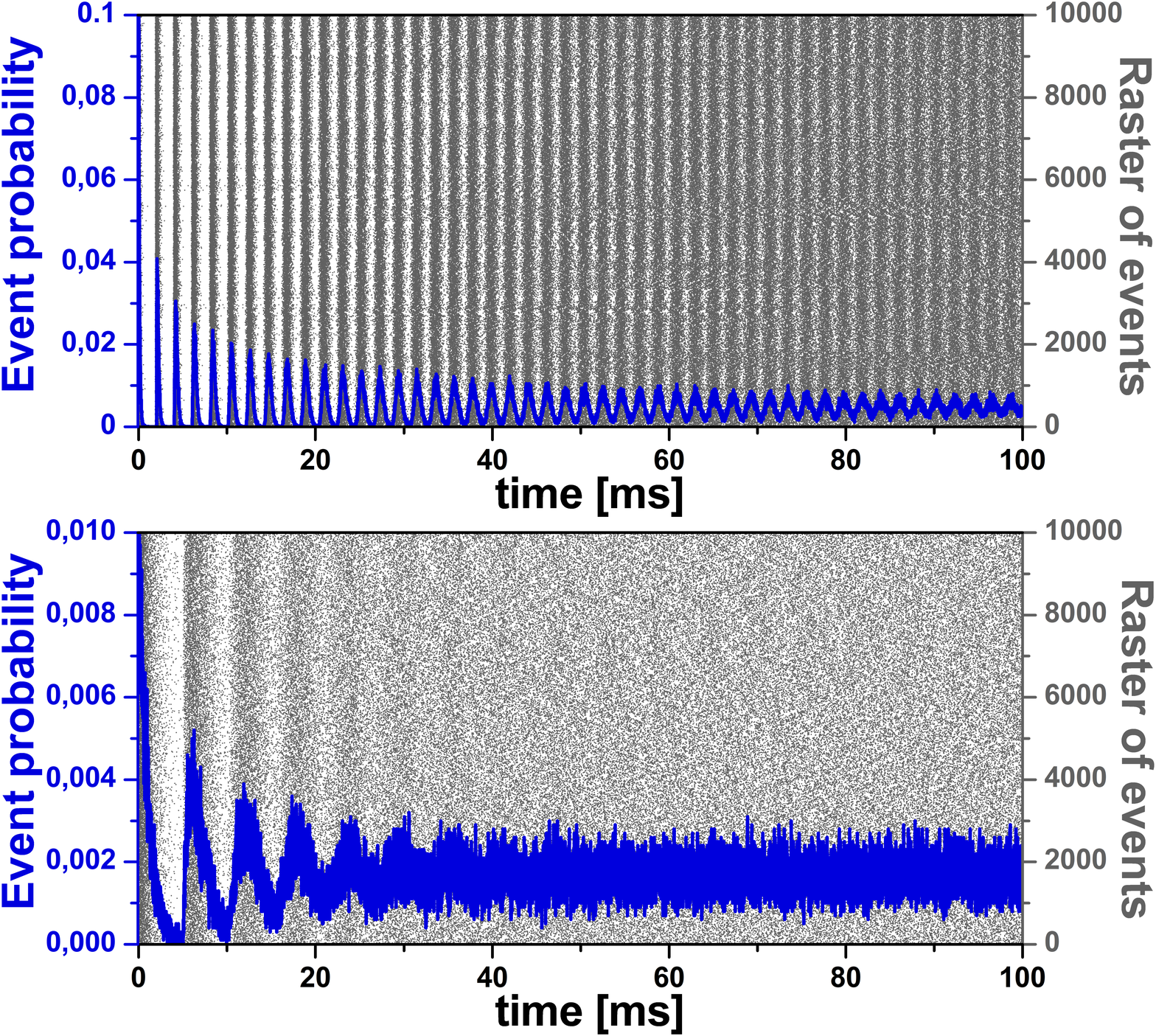} 	
\caption{Numerical simulation of time dependence of the event probability, where an event is spike generation, for $10^{4}$ disconnected stochastically-spiking neurons at time step $\triangle t$ = 0.01 ms. Top: Raster of events (gray dots, scale on the right) and the corresponding time dependence of the event probability (blue line, scale on the left) at $p_{s}$ = 0.1 and $\tau_{ref}$ = 2 ms. It is seen that the period of damped oscillations of the probability is $\tau_{ref}$. The asymptotic value to which the damped oscillations converge corresponds to the calculated value $\bar{p}_{s}=4.76\cdot 10^{-3}$ (or the average event frequency $\nu = 476$ Hz, see Eq. (\ref{nu})). Note that the nonphysiologically large value of $p_{s}$ was taken solely for the illustrative purpose. Bottom: Analogous graphs for $p_{s}$ = 0.01 and $\tau_{ref}$ = 5 ms. In this case, $\bar{p}_{s}=1.67\cdot 10^{-3}$ and $\nu = 167$ Hz.}
\label{Fig1}
\end{figure}

Performing either a large number of repeated passes of the observation interval $T_{obs}$ for a single neuron or a single pass for the large ensemble of independent neurons, one gets the statistical distribution of the event occurrence in the entire sequence of $N$ time intervals. Normalizing this distribution by the number of either the passes or the neurons in the ensemble, one can obtain the probability distribution $P_{k}$ of spike generation at $k$-th elementary time step of the observation interval. Due to the refractory period, the event probability, as a function of time, exhibits damped oscillations with the average period equal to $\tau_{ref}$ (Fig. 1).

In turn, the analytical problem consists in finding the probability $P_{k}$ of spike generation at each $k$-th elementary time step so that in the asymptotic limit $k\rightarrow\infty$ one would obtain $P_{k}\rightarrow\bar{p}_{s}$.

\bigskip

\noindent \textbf{3. Derivation of the exact analytical formula for $P_{k}$}

\bigskip

\noindent The probability of spike generation at the $k$-th time step ($k=\overline{1,N}$), $P_{k}\equiv P(A_{k})$, is equal to the product of $p_{s}$ and the probability that in the interval from $k-n_{ref}$ to $k-1$ inclusively no spike has been emitted,
\begin{equation}
P_{k}=p_{s}\left[  1-P(A_{k-n_{ref}}+A_{k-n_{ref}+1}+...+A_{k-1})\right].\label{1}
\end{equation}
The events of spike generation in $n_{ref}$ consecutive time intervals are pairwise incompatible. Therefore the probability of the sum in Eq. (\ref{1}) equals the sum of probabilities
\begin{equation}
P(A_{k-n_{ref}}+A_{k-n_{ref}+1}+...+A_{k-1})={\displaystyle\sum\limits_{j=k-n_{ref}}^{k-1}}P(A_{j}),\label{2}
\end{equation}
and the sought-for probability at the $k$-th step is determined in a recurrent manner, with the recursion period equal to the refractory period (cf. \cite{JNM01}),
\begin{equation}
P_{k}=p_{s}(1-%
{\displaystyle\sum\limits_{j=k-n_{ref}}^{k-1}}P_{j})=p_{s}\left[  1-(S_{k-1}-S_{k-n_{ref}-1})\right],\label{3}
\end{equation}
where, by definition,
\begin{equation}
S_{m}=
\begin{dcases}
0, \text{  } m \le 0, \\
{\displaystyle\sum\limits_{j=1}^{m}}P_{j}, \text{  } m>0.
\end{dcases}\label{4}
\end{equation}
Such a definition of $S_{m}$ allows us to directly generalize the formula for $P_{k}$ to the range $1\leq k\leq n_{ref}$. The closed formula for $S_{k}$ has the form of a linear recurrent sequence
\begin{equation}
S_{k}=P_{k}+S_{k-1}=p_{s}(1+S_{k-n_{ref}-1})+(1-p_{s})S_{k-1}.\label{5}%
\end{equation}
The resulting formula (\ref{3}) accurately describes the numerical statistics (Fig. 2).
\begin{figure}[!t]
\centering
\includegraphics[width=0.8\textwidth]{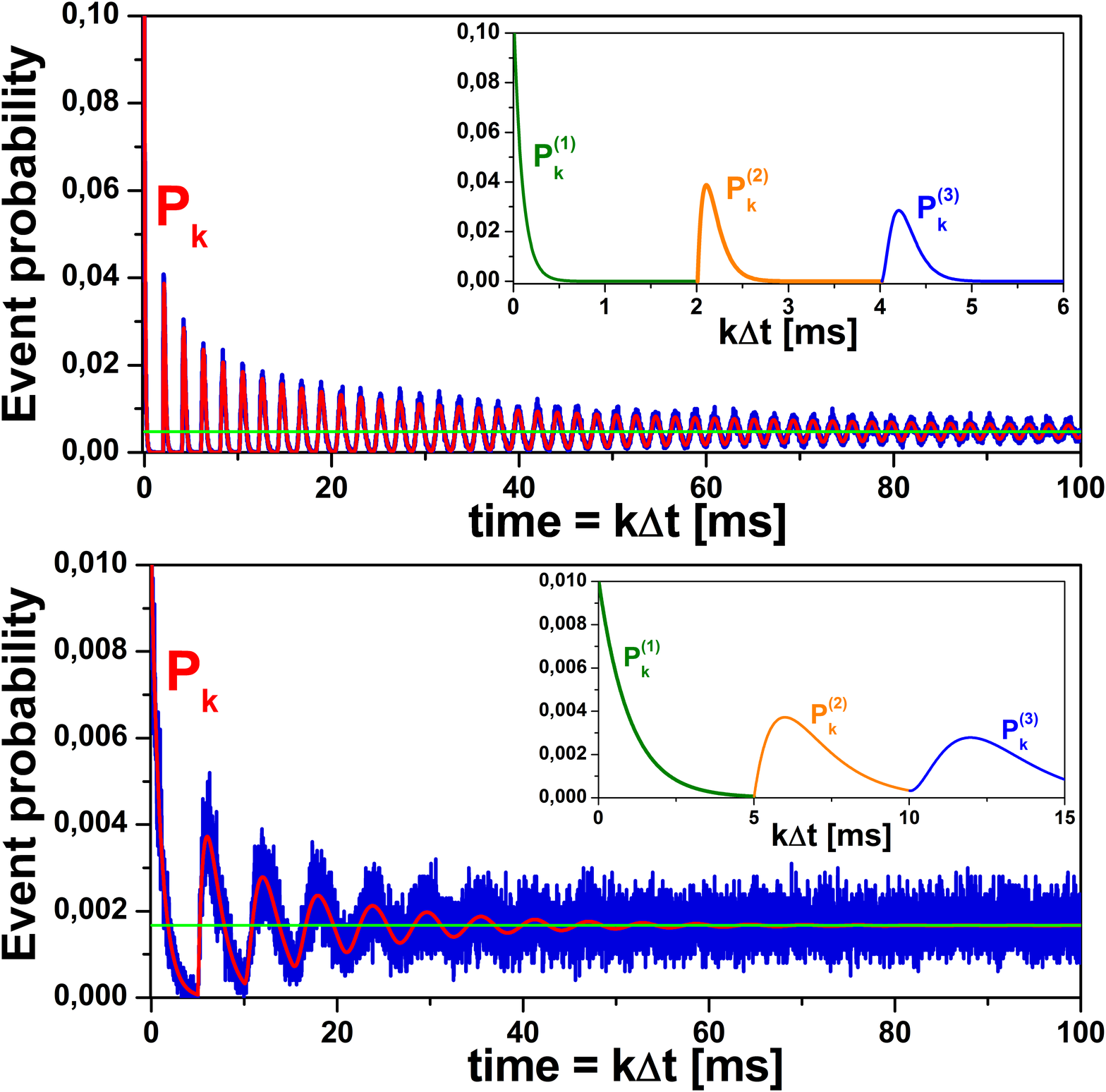} 	
\caption{Comparison of the results of numerical simulations (see Fig. 1) with the calculation by mutually equivalent analytical formulas (\ref{3}), (\ref{Pk}), (\ref{pol}), (\ref{8}), and (\ref{9}). Top: Time dependence of the event probability (blue line, data taken from the top graph in Fig. 1) and the corresponding analytical curve for $P_{k}$ (red line) at $p_{s}$ = 0.1 and $\tau_{ref}$ = 2 ms. The asymptotic value $P_{\infty}=\bar{p}_{s}=4.76\cdot 10^{-3}$ is shown by the green horizontal line. Inset: Partial analytical curves for the first three refractory intervals calculated by formulas (\ref{Pk1}), (\ref{Pk2}), and (\ref{Pk3}). Bottom: Analogous graphs (both the main graph and the inset) for different parameter values, $p_{s}$ = 0.01 and $\tau_{ref}$ = 5 ms, with the asymptotic probability $P_{\infty}=1.67\cdot 10^{-3}$.}
\label{Fig2}
\end{figure}

One should note three important consequences. First, for the range $1\leq k\leq n_{ref}+1$, where $S_{k}=p_{s}+(1-p_{s})S_{k-1}$, the sum $S_{k}$ can be easily found explicitly, as it is an arithmetic-geometric progression of the form $S_{k}=rS_{k-1}+d$ with $S_{1}=p_{s}$, $r=1-p_{s}$, and $d=p_{s}$. According to the formula for the explicit form of the $k$-th term of this progression,
\[
S_{k}=S_{1}r^{k-1}+\frac{r^{k-1}-1}{r-1}d=1-(1-p_{s})^{k}.
\]
Hence, from Eq. (\ref{3}) the probability of spike generation in the interval $1\leq k\leq n_{ref}+1$ is given by
\begin{equation}
P_{k}\equiv P_{k}^{(1)}=p_{s}(1-S_{k-1})=p_{s}(1-p_{s})^{k-1}.\label{Pk1}
\end{equation}
Here and below, the upper index $m$ in parentheses in $P_{k}^{(m)}$ indicates the sequential number of the current refractory interval counted from the initial moment $t=0$. The probability (\ref{Pk1}) refers to the geometric distribution and, given the initial condition, naturally coincides with the probability of the first spike generation at the arbitrary $k$-th time step, such that ${\displaystyle\sum\limits_{k=1}^{\infty}}P_{k}^{(1)}=1$.

The second consequence is that formula (\ref{3}) makes it easy to obtain the asymptotic probability value $P_{k}$ at $k\rightarrow\infty$. Indeed, the difference $S_{k-1}-S_{k-n_{ref}-1}$ contains $n_{ref}$ terms. At $k\rightarrow\infty$ the probability at the $k$-th step remains practically unchanged. Denoting it as $P_{\infty}$, from Eq. (\ref{3}) one gets $P_{\infty}=p_{s}(1-n_{ref}P_{\infty})$, whence
\begin{equation}
P_{\infty}=p_{s}/(1+n_{ref}p_{s}).\label{7}
\end{equation}

Finally, the third consequence: calculating the adjacent terms $P_{k+1}$ or $P_{k-1}$, alike to Eq. (\ref{5}), one can exclude sums (\ref{4}) from Eq. (\ref{3}) and obtain linear recurrent sequence for $P_{k}$:
\begin{equation}
P_{k}=
\begin{dcases}
p_{s}(1-p_{s})^{k-1}, \text{  } k \leq n_{ref}+1, \\
p_{s}P_{k-n_{ref}-1}+(1-p_{s})P_{k-1}, \text{  } k>n_{ref}+1.
\end{dcases}\label{Pk}
\end{equation}
This formula is completely equivalent to Eq. (\ref{3}). Using Eq. (\ref{Pk}), one can easily find the expression for $P_{k}$ in an explicit form within the intervals of $k$ multiples of $n_{ref}+1$.

For example, at $n_{ref}+1 \leq k \leq 2(n_{ref}+1)$ one gets
\begin{equation}
P_{k}^{(2)}=(k-n_{ref}-1)p_{s}^{2}(1-p_{s})^{k-n_{ref}-2} + P_{k}^{(1)}=P_{k}^{(1)}[1+(k-n_{ref}-1)q],\label{Pk2}
\end{equation}
where $q = p_{s}(1-p_{s})^{-(n_{ref}+1)}$.

Next, at $2(n_{ref}+1) \leq k \leq 3(n_{ref}+1)$ one gets
\begin{align}
P_{k}^{(3)} & =\frac{1}{2}(k-2n_{ref}-2)(k-2n_{ref}-1)p_{s}^{3}(1-p_{s})^{k-2n_{ref}-3} + P_{k}^{(2)}=\\
& =P_{k}^{(1)}[1+(k-n_{ref}-1)q+\frac{1}{2}(k-2n_{ref}-2)(k-2n_{ref}-1)q^{2}].\label{Pk3}
\end{align}
It is worth noting that in Eqs. (\ref{Pk2}) and (\ref{Pk3}) the numerical coefficient in the highest-order term with respect to $q$ is the so-called triangular number $j(j+1)/2$ at $j=k-2n_{ref}-2$. The inset in Fig. 2 shows the plots for $P_{k}^{(1)}$, $P_{k}^{(2)}$, and $P_{k}^{(3)}$.

By the induction method, one can obtain a formula for $P_{k}^{(m+1)}$, which is valid within the range $m(n_{ref}+1) \leq k \leq (m+1)(n_{ref}+1)$:
\begin{equation}
P_{k}^{(m+1)}=P_{k}^{(1)}[\frac{1}{m!}\prod\limits_{j=1}^{m}(k-m\cdot n_{ref}-j)]q^{m} + P_{k}^{(m)}=P_{k}^{(1)}C_{k-m\cdot n_{ref}-1}^{m}q^{m} + P_{k}^{(m)}.\label{Pkmp1}
\end{equation}
Here, equality
\begin{equation}
\frac{1}{m!}{\displaystyle\prod\limits_{j=1}^{m}}(n-j)=C_{n-1}^{m},
\end{equation}
where $n>m \geq 1$, and $C_{n}^{k}=\frac{n!}{k!(n-k)!}=\dbinom{n}{k}$ is the standard binomial coefficient, has been used to get the last formula in Eq. (\ref{Pkmp1}).

In turn, Eq. (\ref{Pkmp1}) allows one to derive the general explicit expression for $P_{k}$ in the polynomial form:
\begin{equation}
P_{k}=P_{k}^{(1)}(1+{\displaystyle\sum\limits_{i=1}^{m}}C_{k-i\cdot n_{ref}-1}^{i}q^{i}),\label{pol}
\end{equation}
where $m$ is the integer part of the rational number $k/(n_{ref}+1)$.

Another equivalent formula for $P_{k}$ can be obtained as follows. Denote $P(A_{k}|A_{j})$ the conditional probability of spike generation at the $k$-th step, provided that the previous spike was generated at the $j$-th step. At times greater than the refractory period, i.e. at $k > n_{ref}$, the probability of generating a subsequent spike depends only on the moment of the previous spike. Therefore, taking into account the initial condition, $P(A_{k}|A_{j})=P_{k-j-n_{ref}}^{(1)}$ at $k-j>n_{ref}$, and $P(A_{k}|A_{j})=0$ at $k-j\leq n_{ref}$.

If $k>n_{ref}+1$, probability $P_{k}$ can be written as a sum of two terms: (i) the probability $P_{k}^{(1)}$ that a spike will be emitted for the first time on the $k$-th step and (ii) the probability that at least one spike has been emitted previously. The latter has the form of a convolution and follows from the total probability formula. In particular, one gets
\begin{equation}
P_{k}=P_{k}^{(1)}+\sum\limits_{j=1}^{k-1}P_{j}P(A_{k}|A_{j})=P_{k}^{(1)}+\sum\limits_{j=1}^{k-n_{ref}-1}P_{j}P_{k-j-n_{ref}}^{(1)}. \label{8}
\end{equation}
Notably, using substitution $i=k-j-n_{ref}$, one can virtually swap the indices of the multipliers under the sign of the sum in Eq. (\ref{8}), while the formula does not change its numerical value:
\begin{equation}
P_{k}=P_{k}^{(1)}+\sum\limits_{i=1}^{k-n_{ref}-1}P_{k-i-n_{ref}}P_{i}^{(1)}. \label{9}
\end{equation}
Despite the different appearance in relation to Eqs. (\ref{3}) and (\ref{Pk}), the last two formulas lead to the same numerical results. In fact, these can be derived from the recurrent formula (\ref{Pk}): e.g., Eq. (\ref{9}) can be obtained by successively substituting values $P_{k-1}$, $P_{k-2}$, $\ldots$ , $P_{k-n_{ref}-1}$ into Eq. (\ref{Pk}) and using the definition (\ref{Pk1}) for $P_{k}^{(1)}$.

\bigskip

\noindent \textbf{4. Damping of the oscillations}

\bigskip

\noindent The explicit formulas (\ref{Pk2}) and (\ref{Pk3}) allow analytical calculation of the relative damping of the second and third peaks of $P_{k}$. Finding the location of these peaks corresponds to solving a linear and quadratic algebraic equation, respectively. In particular, for $(n_{ref}+1)\leq k\leq 2(n_{ref}+1)$, assuming that $k$ is a continuous variable and calculating $dP_{k}^{(2)}/dk = 0$ with the use of Eq. (\ref{Pk2}), one gets the location of the second peak,
\begin{equation}
k_{\max2}=n_{ref}+1+R,
\end{equation}
where $R = 1/u-1/q$, $u = \ln[1/(1-p_{s})]$, and $q = p_{s}(1-p_{s})^{-(n_{ref}+1)}$. Substituting $k_{\max2}$ into Eq. (\ref{Pk2}), we get the amplitude of the second peak
\begin{equation}
P_{\max}^{(2)}\equiv P_{k_{\max2}}^{(2)}=P_{k_{\max2}}^{(1)}\frac{q}{u}=\frac{p_{s}^{2}}{u}(1-p_{s})^{R-1}.\label{Pmax2}
\end{equation}
The damping can be defined as the ratio of amplitudes of the adjacent decreasing peaks,
\begin{equation}
D_{i+1}=P_{\max}^{(i+1)}/P_{\max}^{(i)}.
\end{equation}
Given that $k_{\max1} = 1$ and $P_{k_{\max1}}^{(1)}=p_{s}$, for the second peak we get
\begin{equation}
D_{2}=\frac{p_{s}}{u}(1-p_{s})^{R-1}.\label{D2}
\end{equation}
For the third refractory interval, $2(n_{ref}+1)\leq k\leq3(n_{ref}+1)$, calculation of $dP_{k}^{(3)}/dk =0$ using Eq. (\ref{Pk3}) results in quadratic equation,
\begin{equation}
xk^{2}-yk-z=0,\label{quad}
\end{equation}
where $x=\frac{1}{2}q^{2}u$, $y=q^{2}+(2n_{ref}+\frac{3}{2})q^{2}u-qu$, and
\[
z=-(2n_{ref}+\frac{3}{2})q^{2}-(n_{ref}+1)(2n_{ref}+1)q^{2}u+q+(n_{ref}+1)qu-u.
\]
A suitable solution of Eq. (\ref{quad}) is the root
\begin{equation}
k_{\max3}=\frac{y+\sqrt{y^{2}+4xz}}{2x}=2(n_{ref}+1)+R+X,\label{k_max3}
\end{equation}
where
\begin{equation}
X=-\frac{1}{2}+\sqrt{\frac{1}{4}+\frac{1}{u^{2}}-\frac{(2n_{ref}+1)}{q}-\frac{1}{q^{2}}}.\label{X}
\end{equation}
Substituting $k_{\max3}$ into Eq. (\ref{Pk3}) for $P_{k}^{(3)}$ yields
\begin{equation}
P_{k_{\max3}}^{(3)}=P_{k_{\max3}}^{(1)}[1+(n_{ref}+1+R+X)q+\frac{1}{2}(R+X)(1+R+X)q^{2}].
\end{equation}
After elementary but cumbersome calculations (given in the Supplementary Material), one can get a compact analytical expression for the damping coefficient of the third peak,
\begin{equation}
D_{3}=P_{k_{\max3}}^{(3)}/P_{k_{\max2}}^{(2)}=p_{s}(\frac{1}{2}+X+\frac{1}{u})(1-p_{s})^{X}.\label{D3}
\end{equation}
Numerical calculations have confirmed the validity of the obtained formulas. Below we have also listed the numerical values of the relevant quantities, calculated by the above formulas, for two examples shown in Fig. 1 and Fig. 2.

For the first example, at $\tau_{ref}=2$ ms ($n_{ref}=\tau_{ref}/\triangle t=200$ at $\triangle t=0.01$ ms) and $p_{s}=0.1$, we get $q \approx 1.6\times10^{8}$, $u \approx 0.1$, $R \approx 9.5$, $k_{\max2} = 210$, $P_{k_{\max2}}^{(2)} \approx 0.04$, $k_{\max3} = 420$, $P_{k_{\max3}}^{(3)} \approx 0.03$, $D_{2} \approx 0.39$, and $D_{3} \approx 0.74$.

For the second example, at $\tau_{ref}=5$ ms ($n_{ref}=500$) and $p_{s}=0.01$, we get $q \approx 1.5$, $u \approx 0.01$, $R \approx 99$, $k_{\max2} = 599$, $P_{k_{\max2}}^{(2)} \approx 0.004$, $k_{\max3} = 1196$, $P_{k_{\max3}}^{(3)} \approx 0.003$, $D_{2} \approx 0.37$, and $D_{3} \approx 0.75$.

It is seen that the numerical values $D_{2} \approx 0.4$ and $D_{3} \approx 0.75$ are fairly robust against changing the parameters. Thus, the magnitude of the second peak is approximately equal to 40\% of the magnitude of the first. In turn, the value of the third peak is approximately 75\% of the value of the second or 30\% of the value of the first peak.

\bigskip

\noindent \textbf{5. Comparison with the renewal theory}

\bigskip

\noindent In the framework of the renewal theory \cite{Cox62,Perk67}, when the time intervals between events are independent random variables, one can standardly find the time dependence of the event probability, provided that an event occurred at the initial moment. In particular, the sought-for probability is expressed through the so-called renewal density $h(t)$,
\begin{equation}
P^{(rd)}_{k}=h(k\triangle t)\triangle t.\label{h1}
\end{equation}
In turn, the renewal density $h(t)$ is determined by the distribution density $f(\tau)$ for the time interval $\tau$ between the successive events \cite{Cox62,Perk67},
\begin{equation}
h(t)={\displaystyle\sum\limits_{n=1}^{\infty}}f_{n}(t),\label{h2}
\end{equation}
where $f_{1}(t)\equiv f(t)$ and for $n\geq2$ functions $f_{n}(t)$ are given by the recursive convolution
\begin{equation}
f_{n}(t)={\displaystyle\int\limits_{0}^{+\infty}}f_{n-1}(x)f(t-x)dx.\label{h4}
\end{equation}
Qualitatively, functions $f_{n}(t)$ are the distribution densities of so-called $n$-th order time intervals between events \cite{Perk67}: the first-order interval is the time elapsed from some benchmark event to the first following event, the second-order interval is the time elapsed from the benchmark event to the second following event, etc. The $n$-th order interval is therefore a sum of $n$ successive first-order intervals and is spanned by $(n + 1)$ successive events.

At asymptotically large time $t\rightarrow+\infty$, $h(t)$ is saturated \cite{Cox62},
\begin{equation}
\lim_{t\rightarrow+\infty}h(t)=\rho.\label{hlim}
\end{equation}
Here $\rho$ is the mean rate of events, defined as the inverse mean interval between the successive events,
\begin{equation}
\rho^{-1}={\displaystyle\int\limits_{0}^{+\infty}}\tau f(\tau)d\tau.\label{h5}
\end{equation}

There are many, likely independent, examples of applying the renewal theory results to the case, where the event occurrence is the Poisson process modulated by the constant dead time (in our notations, absolute refractory period $\tau_{ref}$) or, equivalently, the distribution density of time intervals between the successive events has a form of the shifted exponential distribution,
\begin{equation}
f(t)=\nu_{0}\exp(-\nu_{0}(t-\tau_{ref}))\theta(t-\tau_{ref}),\label{h6}
\end{equation}
where $\nu_{0}=p_{s}/\triangle t$ and $\theta(\ldots)$ is the unit step function: $\theta(x) = 1$ at $x \geq 0$ and $\theta(x) = 0$ otherwise. In particular, to the best of our knowledge, the first relevant paper dates back to 1947 \cite{Malm47} and has been followed by both in-depth mathematical studies \cite{Cox62,Ric66,Muller73,Muller74,Cant75,Pom99,Picib08,PRE2010} and applied studies for neuroscience \cite{Perk67,John83,JCN2012,NC2018,TMC78,JASA85,John86,Koch93,Berry98,Reich98,Kas01} (see also overviews in \cite{John96} and \cite{Gerst02}). Below, we briefly outline and compare the previous results with our findings.
\begin{figure}[!t]
\centering
\includegraphics[width=0.8\textwidth]{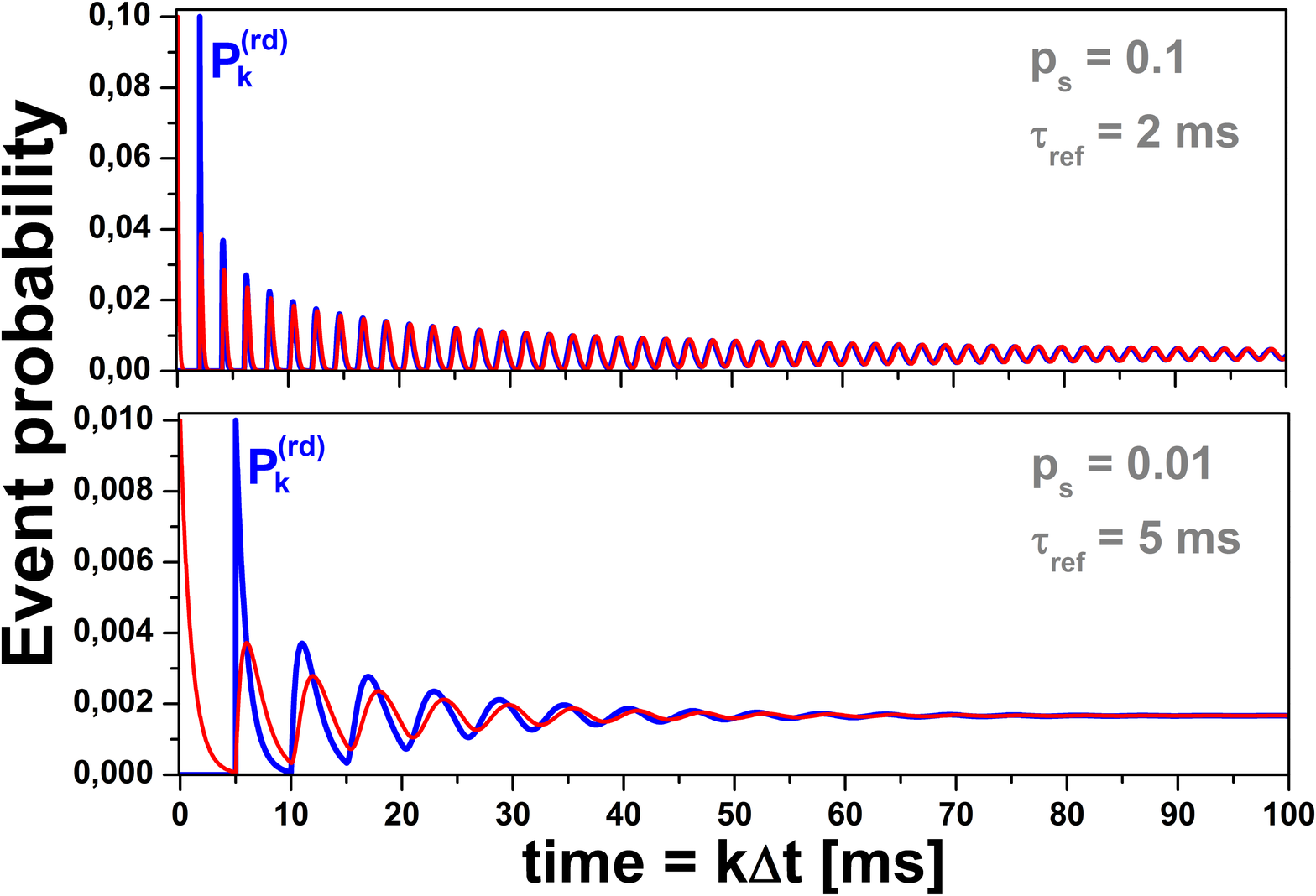} 	
\caption{Comparison of the time dependencies for the event probability $P_{k}$ calculated by mutually equivalent Eqs. (\ref{3}), (\ref{Pk}), (\ref{pol}), (\ref{8}), (\ref{9}) in Section 3 (red curves) and the event probability $P^{(rd)}_{k}$ calculated within the renewal theory approach by Eq. (\ref{h1}) in Section 5 (blue curves). The red and blue curves, accurate to the offset equal to the refractory period, completely coincide. The offset arises due to the different initial conditions (exit from the refractory period for the red curves and spike generation for the blue ones) and has been left intentionally in order to make the curves distinguishable (for the same initial condition, the red and blue curves match accurately). The top plot is for $p_{s}$ = 0.1 and $\tau_{ref}$ = 2 ms. The bottom plot is for $p_{s}$ = 0.01 and $\tau_{ref}$ = 5 ms.}
\label{Fig3}
\end{figure}

For $f(t)$ given by (\ref{h6}), using the Laplace transform, one can reduce Eq. (\ref{h4}) to the following expression
\begin{equation}
f_{n}(t)=\nu_{0}\frac{(\nu_{0}(t-n\tau_{ref}))^{n-1}}{(n-1)!}\exp(-\nu_{0}(t-n\tau_{ref}))\theta(t-n\tau_{ref}),\label{h7}
\end{equation}
which is the probability density function for the Erlang (or gamma) distribution, enabling computation of the renewal density $h(t)$ and the sought-for probability (\ref{h1}). The plot of the function $P^{(rd)}_{k}$ given by Eq. (\ref{h1}) with
\begin{equation}
h(t)={\displaystyle\sum\limits_{n=1}^{n_{\max}}}f_{n}(t),\label{h8}
\end{equation}
where $n_{\max}=10^{3}$ and $f_{n}(t)$ is determined by (\ref{h7}), is shown in Fig. 3. The time dependence $P^{(rd)}_{k}$, accurate to an offset equal to the refractory period, completely coincides with $P_{k}$ calculated by mutually equivalent formulas (\ref{3}), (\ref{Pk}), (\ref{pol}), (\ref{8}), (\ref{9}) in Section 3. The offset arises due to the different initial conditions: an exit from the refractory period at $t=0$ in our model and an event occurrence at $t=0$ in the standard renewal-density approach.

Finally, using (\ref{hlim}), (\ref{h5}) and (\ref{h6}), one can get the asymptotic value of $h(t)$,
\begin{equation}
\lim_{t\rightarrow+\infty}h(t)=\nu_{0}/(1+\nu_{0}\tau_{ref}),\label{h9}
\end{equation}
which is consistent with formulas (\ref{nu}) and (\ref{7}) for $\nu$ and $P_{\infty}$, respectively.

\bigskip

\newpage

\noindent \textbf{6. Discussion}


\noindent In most cases of multi-unit systems, such as biological neuronal networks with neurons as the units, the absolute refractory period $\tau_{ref}$ is not exactly the same for each unit. For instance, it may be randomly distributed so that each neuron has its own value of $\tau_{ref}$ (e.g., see Fig. 2a in \cite{Avi13}). Alternatively, $\tau_{ref}$ may be updated randomly after every event \cite{Pet21}. Then increasing the variance of the refractory period distribution would apparently blur the damped oscillations. Nevertheless, one can safely set the absolute refractory period as a positive constant value, same for all neurons, in the majority of neuronal network models without losing their main predictions. Therefore the obtained analytical results may be directly used as a simple test of correctness for the neuronal network models with stochastic spontaneous activity of single neurons, provided that the stochasticity mimics internal neuronal dynamics rather than incoming synaptic currents from some "external" neurons (e.g., see \cite{PRL93,PRE95,PRE98}, cf. \cite{Newh10}). Indeed, in the former case, one can approximate the probability of spontaneous spike generation as nearly time-independent, constant quantity. On the contrary, in the latter case, this probability may be essentially time-dependent, even for cortical neurons in the high-input regime, when they receive hundreds of excitatory synaptic inputs during each interspike interval \cite{SN1998,NRN2003,Kuhn2004}. Nevertheless, the obtained results may be still applicable for the stochastic firing-rate-based neuron models (e.g., \cite{JN2008,SRM1,SRM2}), where the neuronal dynamics are entirely described by a time-dependent probability of spike generation. It should also be mentioned that there exists a different type of stochasticity induced by the multilayer network structure \cite{Perc16,Perc17} and the results obtained in this paper are not directly applicable to this type.

The test mentioned above is as follows: in the limiting case where all synaptic interactions between neurons are set to zero, one should increase the value of probability of spontaneous spiking by a single neuron (implying that this value has been made the same for all neurons of the network) until the pseudo-network activity (or, more precisely, the fraction of neurons emitted spikes during the elementary time step of simulation dynamics) would exhibit pronounced damped oscillations. Next, turning to the analytical formulas, one should either compute the time dependence of spiking probability (e.g., using Eq. (\ref{Pk})), and match this with the simulation result or, even simpler, compare the relative values of the first, second, and third peaks of the normalized collective spiking activity of neurons with the values obtained by formulas (\ref{D2}) and (\ref{D3}). If the neuronal network model has been designed and implemented correctly, one gets the complete correspondence, as disconnecting neurons of the network turns it into a statistical ensemble of independent neurons (cf. \cite{Truc2017}). In fact, the plots in Fig. 1 have been originated from two actual realizations of the test applied to the spiking neuronal network model reported in \cite{ZP2021}.

In addition, the analytical results may be of interest for developing and testing (i) population-level "neural mass" models and (ii) network-based "cellular automata" models.

The top-down population models, such as the classical Wilson-Cowan model \cite{WCBJ72}, usually include the absolute refractory period as a phenomenological parameter. Notably, the damped oscillations due to the absolute refractory period were the first consequence from the original Wilson-Cowan equations compared to their temporally coarse-grained version (see Fig. 3 in \cite{WCBJ72}), though after some reasoning the authors concluded that "the damped oscillation is not likely to have any functional significance" (\cite{WCBJ72}, page 9). In a more recent bottom-up approach to neuronal population modeling, sometimes refereed as the refractory density approach \cite{Chizh1,Chizh2,Chizh3} (cf. \cite{Omur2000,Gerst2000,Mat2002,Deco2008,Non2010,Gerst17,Dum17,PRE20}), enabling to extract the population dynamics starting with relatively complex, biophysical Hodgkin-Huxley-type neuron models, an occurrence of the damped oscillations due to the refractoriness is the foundation for a standard test on the transient response of the population firing rate to a step-like increase of the common input stimulation.

In turn, a wide spectrum of network-based cellular automata models, which typically constitute randomly-connected simple excitable units with nonlinear interaction, has been elaborated for large-scale numerical and/or analytically-tractable studies of critical collective behavior via avalanches, emergent synchronization and global oscillations of neuronal activity \cite{JN2008,SRM1,SRM2,SR2017} (some earlier findings were reviewed in \cite{Lind2004}). In addition to the foregoing models, some relevant models, where interference between the absolute refractory period and stochastic event occurrence may be essential, are the two-state unit model \cite{Esc2012}, the three-state unit model \cite{PA2003,Wood1,Wood2} (see also \cite{JSM2011,Lima19}), the DeVille-Peskin multi-state model \cite{BMB2008}, the cortical branching model \cite{Widom2014}, and others \cite{Golt10,Lee2014,NJP15}.

Finally, the analytical results obtained in this paper might be useful for a qualitative analysis of statistical data on both spontaneous and stimulated spiking activity of real biological neurons. For instance, the neurons could have receptive fields subject to an intense step-like stimulus (for the fly's photoreceptors, such a case was considered in \cite{Song17}, see Figs. 5 and 6 there). The references to relevant experimental data are as follows: Figs. 6 and 15 in \cite{Pog64}, Fig. 4 in \cite{Rod67}, Fig. 1 in \cite{Gray67},  Figs. 1, 3, 4, 5, and 6 in \cite{Moor70}, Fig. 3 in \cite{Rob87}, Fig. 2B in \cite{Nin95}, Figs. 2H and 8E in \cite{Kita04}, Figs. 4A and 4B in \cite{Kita06}, Fig. 2A in \cite{Beat12}, Fig. 4A in \cite{Lee14}, and Fig. 4A in \cite{Amit17}. In turn, direct quantitative comparison of the analytical results with the above experimental findings is hindered by such additional factors as (i) the relative refractory period, which may be time-dependent, and (ii) uncontrollable non-random interaction of the target neurons with surrounding cells \cite{Kara2000,Kass04,Amar06,Naw07,Maim09,EBR11,Coh11}. These and other factors \cite{Truc05} may have a dominant influence on the heights, widths, and locations of experimental autocorrelogram peaks.

\bigskip

\noindent \textbf{7. Conclusion}

\bigskip

\noindent The model considered in this paper is the Bernoulli scheme supplemented by the condition of absolute refractoriness followed after each random event. It formally refers to the renewal theory. However, being quite simple, the model allows obtaining useful results without invoking this formalism. In particular, four equivalent analytical descriptions of the damped oscillations of the event probability have been given: (i) recurrent formula (\ref{3}) through the difference of two sums, (ii) closed recurrent formula (\ref{Pk}), (iii) explicit formula (\ref{pol}) in the form of a polynomial, and (iv) recurrent convolution-type formulas (\ref{8}), (\ref{9}). It has also been shown that for the Poisson approximation these results accurately coincide with that of the renewal theory. Finally, using the closed recurrent formula, the relative damping coefficients for the second and third peaks of the event probability have been found in the exact analytical form.

\noindent It is worth emphasizing that the example of stochastic neuron spiking serves only as a baseline illustration of the obtained mathematical results, which have potentially much wider range of applicability. In general, the damped oscillations studied in this paper are well-pronounced if the absolute refractory period is a dominantly large timescale in the system's dynamics responsible for the event occurrence.

\bigskip

\noindent \textbf{Data and code availability}

\bigskip

\noindent The Supplementary Material to this paper contains ready-to-use MATLAB/Octave codes for performing simulations and plotting graphs of the obtained analytical formulas. It also contains a detailed derivation of the damping coefficients for the 2nd and 3rd peaks of the event probability (see Section 4).

\bigskip

\noindent \textbf{Author contributions}

\bigskip

\noindent A.V. Paraskevov conceived the study, designed the mathematical model and program codes, performed simulations and analytical calculations, prepared graphs, and wrote the manuscript. A.S. Minkin derived the formulas (\ref{8}) and (\ref{9}), independently reproduced all the rest results, and verified the manuscript.

\bigskip

\noindent \textbf{Declaration of competing interests}

\bigskip

\noindent The authors declare no competing interests.

\bigskip

\noindent \textbf{Acknowledgments}

\bigskip

\noindent A.V. Paraskevov thanks E.Z. Meilikhov, L. Logiaco, and D. Festa for the stimulating discussions.

\end{document}